# Nematic liquid crystals in contact with geometrically and chemically patterned substrates: A Monte Carlo study

D. Jayasri[1*], Regina Jose[2], K. P. N. Murthy[2] and V. S. S. Sastry[2]
[1] Faculty of Mathematics and Physics, University of Ljubljana, 1000 Ljubljana, Slovenia.
[2]School of Physics, University of Hyderabad, Hyderabad 500 046, India.

[*]Corresponding author email address: d.jayasri@fmf.uni-lj.si

## Abstract
Nematic liquid crystals confined to geometrically as well as chemically patterned substrate on one end and a flat substrate with strong anchoring on the other is studied using non-Boltzmann Monte Carlo methods. We observe significant deviations from the continuum-based predictions of the phase diagram which was studied as a function of tilt angle at the top substrate and thickness of the cell. Onset of biaxiality at larger tilt angles at the top substrate is observed. A phase shift introduced between the geometrical and chemical pattern has significant effect on the director structures in the system.

## Introduction
Nematic liquid crystals in contact with solid substrates are extensively used in display devices, like the twisted nematic configurations between flat substrate surfaces. The application of small external field is sufficient to change qualtatively the equilibrium director configurations dictated by the substrates, and thus leading to dramatic changes in the optical properties of the nematic cell. But several inhomogenities do occur naturally as a result of surface treatments like rubbing, needed to induce necessary boundary conditions. In most of the cases such inhomogenities do not reflect in the bulk properties of the nematic cell as the length scales of such patterns are very small compared to the thickness of the cell and also to the wavelength of the visible light. More recent developments however have demonstrated that surfaces patterned with large periodicity are of considerable interest from technological point of view like, for example, in flat panel displays with wide viewing angles ( [1] and references therein). In view of their potential technological importance, it is essential, from the basic science point of view to understand the anchoring effects in detail, and also investigate possible phase transitions between various nematic textures induced by non-uniform surface interactions. Moreover, these systems could prove to be more curious if they are found to exhibit bistable nematic states with two different director orientations possible at same energy [2-6]. Whereas many experimental and theoretical studies have concentrated on either geometrically [7-13,26] or chemically patterned surfaces [14-19], nematic liquid crystals confined to both geometrically structured and chemically patterned substrates have been objects of investigation only recently [17, 20–22]. For example, studies based on the continuum theory of nematic liquid crystals in contact with certain specific geometrically and chemically patterned substrates predicted possible transition between two nematic textures, the so-called H and HAN phases, on varying the thickness of the cell or changing the anchoring angle, $\theta_D$ at the top substrate as depicted in figure 1 [17, 20, 22]. In such studies, the interaction of the nematic liquid crystal system in contact with sinusoidal grating with alternating patterns of homeotropic and planar anchoring is accounted for, based on Frank-Oseen model for distortion free energy [23, 24], while the surface energy function is written in terms of Rapini-Papoular expression [25].

The theoretical treatments normally proceed by reducing the dimensionality of the system for convenience of analysis by including translational periodicity along one of the directions, and sometimes by performing a conformal mapping appropriately to eliminate one more dimension. Further a planar surface inducing an effective anchoring angle is assumed at the grooved surface mimicking the geometrical pattern [17, 20, 22, 26]. The underlying argument relies on the observation that the effect of the patterned surface does not extend into the film beyond the length scale of the geometric structure and hence can be replaced by an effective free energy expansion. In all the cases the equilibrium configurations under different distorting conditions are obtained as corresponding to the minimum values of the effective free energy so proposed.

One of the reasons for attempting these simplifications has been the prohibitive effort involved in numerically tackling the problem within continuum approximation. This extremization procedure within this limit inter alia does not allow for possible thermal effects which could be important in principle in real nematic samples. It could be also interesting from the point of our understanding the thermal effects on these systems to follow the director distributions as the system is cooled gradually from the isotropic phase. Keeping these curiosities in view, we perform Monte Carlo simulations of the model liquid crystal system with interactions among the molecules given by Lebwohl-Lasher potential [27] and impose the boundary conditions to mimic geometrical as well as chemical patterns. Keeping in view the suggestions in the earlier work that determining free energy barriers and detecting metastable states are crucial to investigate such nematic devices (which are not possble to obtain via canonical sampling MC methods) [28], we apply here entropic sampling methods [29-32] in our work.

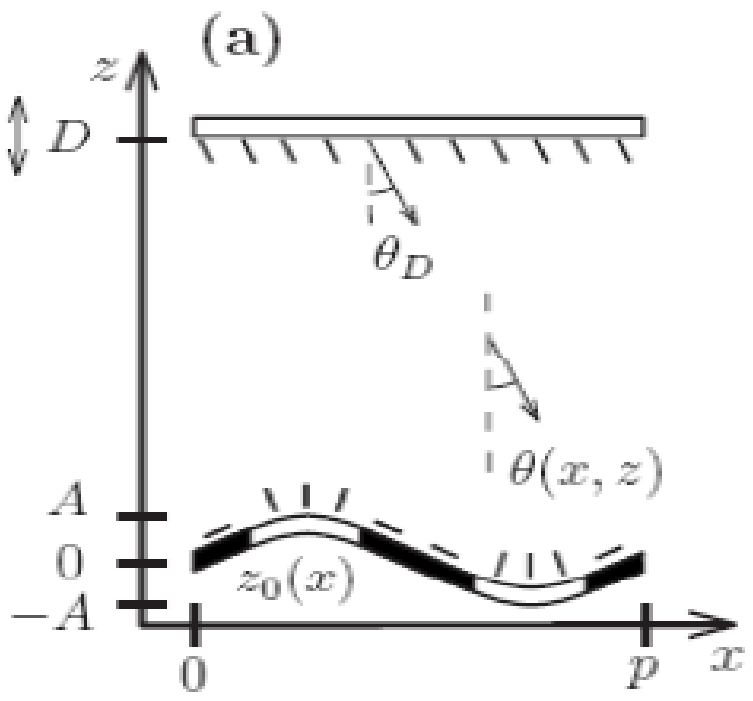

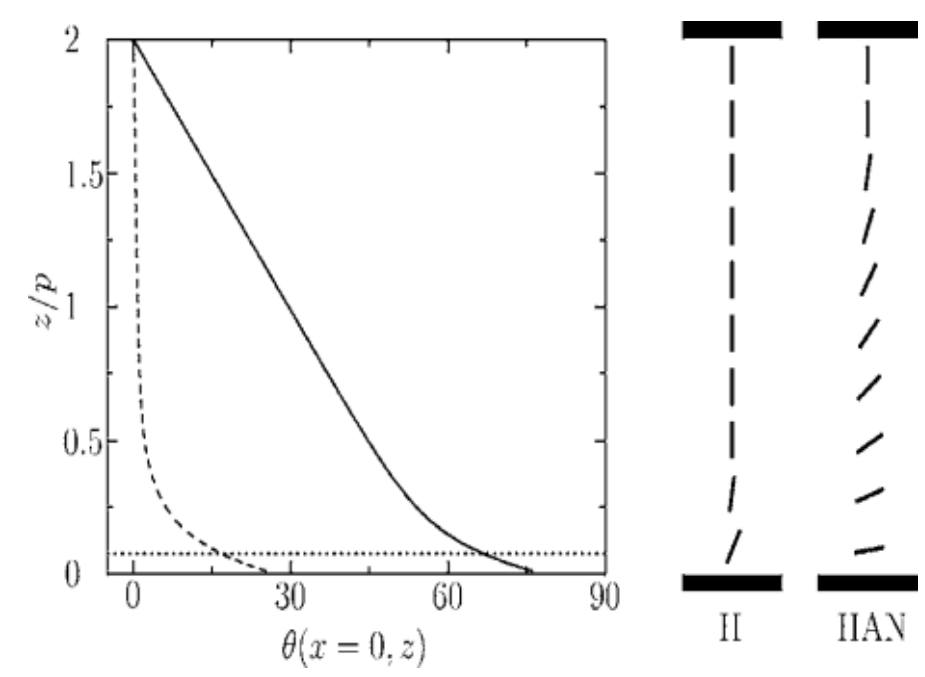

Figure 1: Model liquid crystal cell confined to a sinusoidal substrate with chemical pattern on it (left) [8]. Figure 2: Nematic director structures obtained by changing

the angle, $\theta_D$ (right) [5].

In this letter, we choose a specific geometerical and chemical pattern which has been recently analysed analytically via free energy mimization procedure, and study the effect of different control parameters on the formation of the film as the temperature is cooled, and investigate the stable structures at very low temperatures. We compute the development of free energy surfaces and correlate it with the distribution of different microstates of different types of order but with the same energy, collected in the entropic ensemble. The paper is organized as follows: Section-I deals with the model and the simulation methodology followed; Section II describes the results and discussion and relevant conclusions are drawn in Section III.

## I. The Model and Simulation methodology

Lebwohl – Lasher potential [28] describing the interaction between two nearest neighbouring liquid crystal molecules placed on a cubic lattice cell is given by:

$$H = -\sum_{i,j} \epsilon_{ij} P_2(\cos\theta_{ij}) \qquad (1)$$

Here, $\epsilon_{ij} = \epsilon$ if two nearest neighbours are liquid crystal molecules and 0 otherwise; $\epsilon_{ij} = \omega$ anchoring strength, if one of the nearest neighbours is substrate molecule, ω varies between 0 and 1.

The top layer is a solid substrate inducing strong homeotropic anchoring at an angle, $\theta_D$ as shown in the figure 2. A sinusoidal pattern is carved out of the bottom layers (say, with amplitude, A) along the x-axis. The thickness of the cell, D along the z-direction is taken to be large compared to the amplitude of the sinusoidal pattern. The surface induced interaction is invariant along the y-direction. These results are obtained under the simplifying condition that the cell thickness is much larger than the period of the sinusoidal grating at the bottom surface. Further, the amplitude of the sinusoidal geometric structure is supposed to be a very small fraction of its wavelength. Now, in order to mimic this scenario on a lattice model, we choose a cubic lattice of size 16 × 4 × 66, in the notation of x, y and z dimensions shown in the figure 1. We introduce a sine wave extending over 16 lattice points in the x-dimension, and a film thickness (z-direction) of 66 units. While the y-dimension was argued to be unimportant since the problem under simplifying assumptions is reducible to x-z plane, we extend the lattice along y-direction also over 4 units, in order to facilitate the interactions among different mesogenic units to take place in 3-dimensional space. Periodic boundary conditions are applied along x and y directions. The amplitude of the sinusoidal grating was initially chosen to be 2 units. While such a choice does not implement the geometrical constraints implied in figure 1 accurately due to the limitations introduced by the discretization of space into lattice points, it does capture essential features of the underlying model, the assumption in the continuum treatment being that any arbitrary distribution over the grating period can always be effectively taken into account by introducing an effective tilt angle (say, $\theta_D$) for purposes of predicting the bulk behaviour of the film. The focus of this simulation is thus more on the process of formation of the director structures as nematic phase forms, and on the realizability of H and HAN structures at low enough temperatures, as were predicted by continuum models.

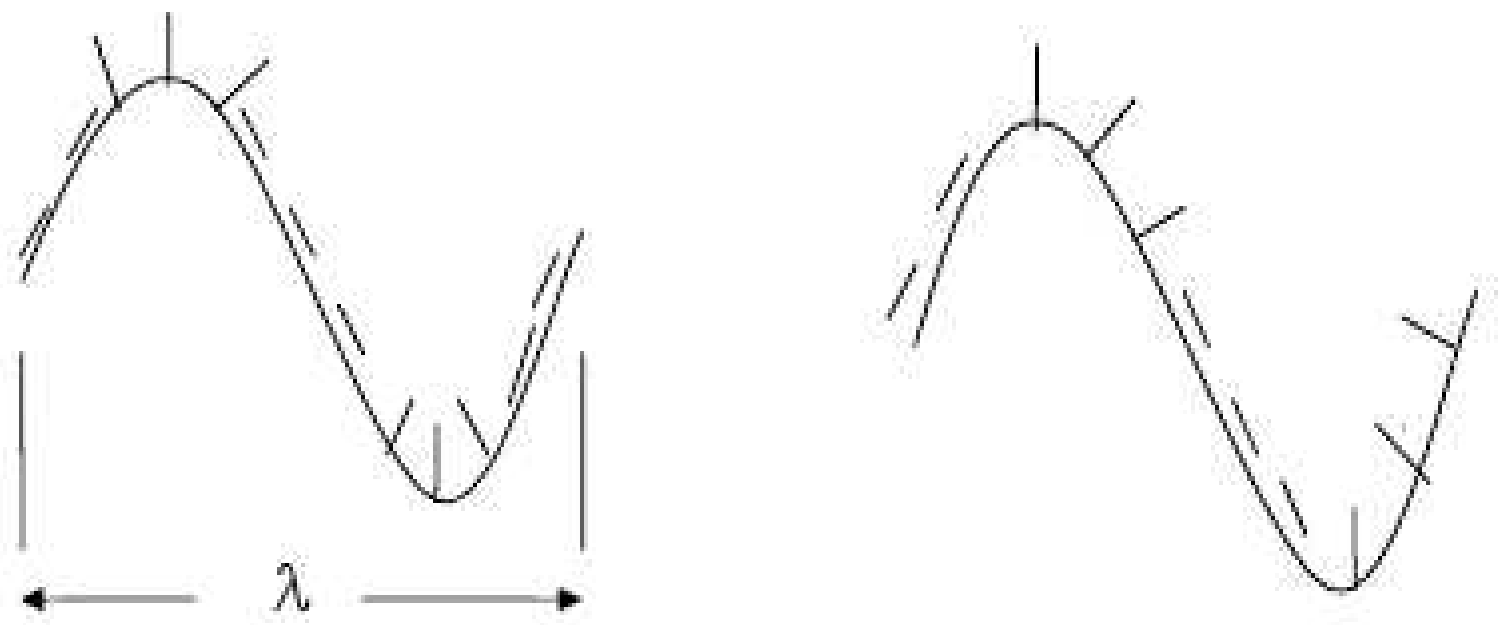

Figure 2: Model A (left): Chemical pattern is in phase with the geometrical pattern; Model B (right): chemical pattern is shifted by $\lambda/8$ with respect to the geometrical pattern.

We consider two model systems in the present work according to the phase shift introduced between the chemical and geometrical patterns on the bottom substrate, see figure 5.2. In Model A, the chemical pattern is in certain phase with respect to the geometrical pattern. Homeotropic alignment is induced at crests and troughs of the sinusoidal wave each for a period of one fourth of the period, while in the other regions locally planar alignment is imposed. In Model B, the chemical pattern is shifted by one-eighth of the wavelength with the geometrical pattern relative to Model A. The anchoring directions in the two models are shown in the figure 5.2.

We employ the modified frontier sampling method [32] to study the system for different angles, $\theta_D$ (between $0^0$ and $45^0$) in the nematic region. We compute the density of states and hence free energy profiles of the two systems. From the entropic ensemble of states, we obtain the canonical ensemble of states at different temperatures. Thus for each value of $\theta_D$, we compute equilibrium averages of different relevant physcial observables as a function of temperature. We also examine the distribution of different regions of energy and look for correlations of these features with corresponding regions in the free energy profiles.

**II. Results and Discussion**

The relevant physical properties, that describe adequately the symmetry of the director field arising from orientational alignment of uniaxial molecules, are the extent of uniaxial order and the possible loss of cylindrical symmetry around the principal director represented by the biaxiality parameter. Figure 3 shows the biaxiality in the system which persists even at the low temperatures as we increase the angle $\theta_D$ from $0^0$ to $45^0$. For Model B, we observe that the biaxiality is present in the system even when the anchoring is perfectly homeotropic at the top substrate. It is interesting to note that where as uniaxial order of the system does not show reasonable change as we increase the angle $\theta_D$ from 0 to 45, biaxiality of the system increases by considerable amount.

Figure 4 depicts the distribution of microstates contained in the entropic ensemble in Model A and Model B (at $\theta_D = 0^0$) with respect to their energy sorted as per their uniaxial and phase biaxial order parameter values. Figure 5 presents these distributions for the case of $\theta_D = 45^0$. Noting that the extent of dstribution of these states over the corresponding order parameters at a given energy (bin) value is a measure of their susceptibility, it may be noted that in Model A homeotropic anchoring at the top substrate ( $\theta_D = 0^0$) has a tendency to form a fairly pure uniaxial phase: the fluctuations in this case with respect to uniaxial order diminish noticeably at low temperatures while there is no significant distribution of microstates in this system with respect to biaxiality parameter (figure 4). In contrast, for the same Model at $\theta_D = 45^0$, figure 5, the uniaxial order has higher degree of fluctuations, while permitting microstates distributed, somewhat widely, over a finite range of biaxiality parameter. This suggests the onset of a small biaxiality with relatively shallow free energy profiles allowing for larger excursions in the order parameters.

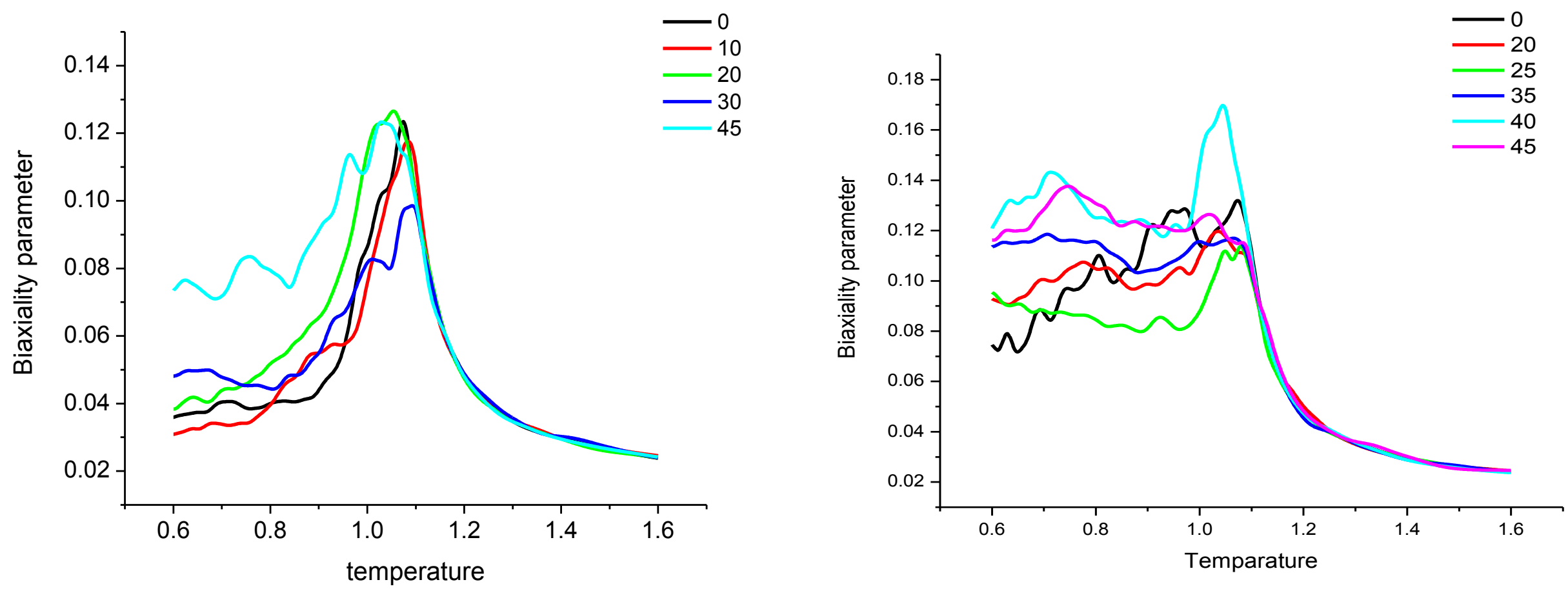

*Figure 3: Biaxiality parameters obtained for different angles, $\theta_D$ ($0^0$ to $45^0$) at the top substrate for Model A (left) and Model B (right) respectively.*

In Model B, on the other hand, similar scenario (corresponding to Model A at $\theta_D = 45^0$ already exists at $\theta_D = 0^0$ (figures 4 and 5). With increase of $\theta_D$ to $45^0$, this model acquires a much wider distribution of microstates with respect to both the order parameters figure 5), and more prominently the biaxiality parameters has a higher average value with longer range of fluctuations. These observations are of course borne out by the corresponding derived variables (figure 3).

Figures 6 and 7 represent the stacked plots of free energy profiles as a function of temperature and order parameter obtained from the entropy generated from the simulations. At lower temperatures we observe significant fluctuations on the surfaces (more clearly in Model B). On close observation, we find that each such fluctuation is a local minimum which exactly matches with the branches in the microstate space we observed in the figures 4 and 5.

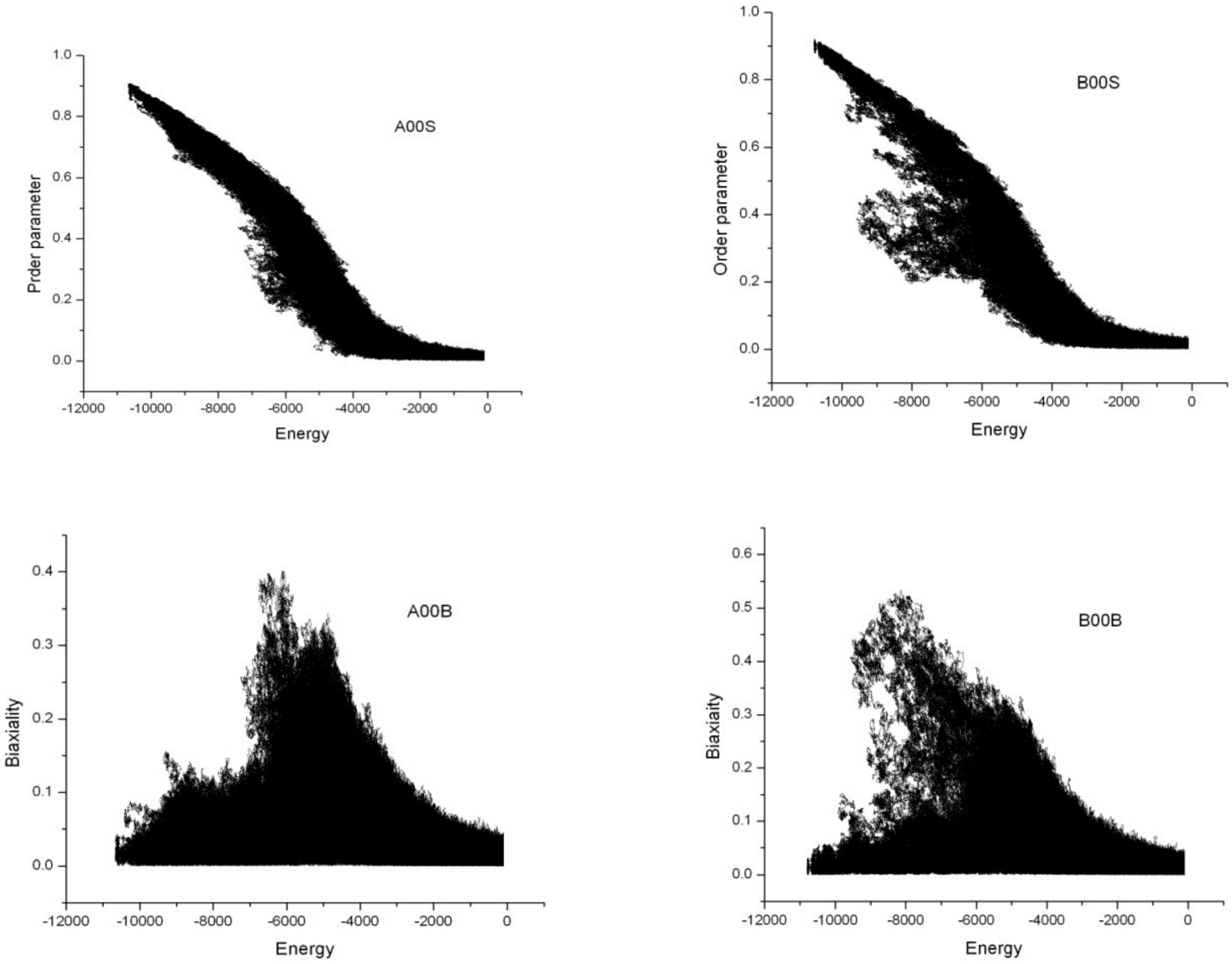

Figure 4: Uniaxial (top) and biaxial (bottom) order parameters for $\theta_D = 0^0$ (left) and $45^0$ (right) respectively for Model A.

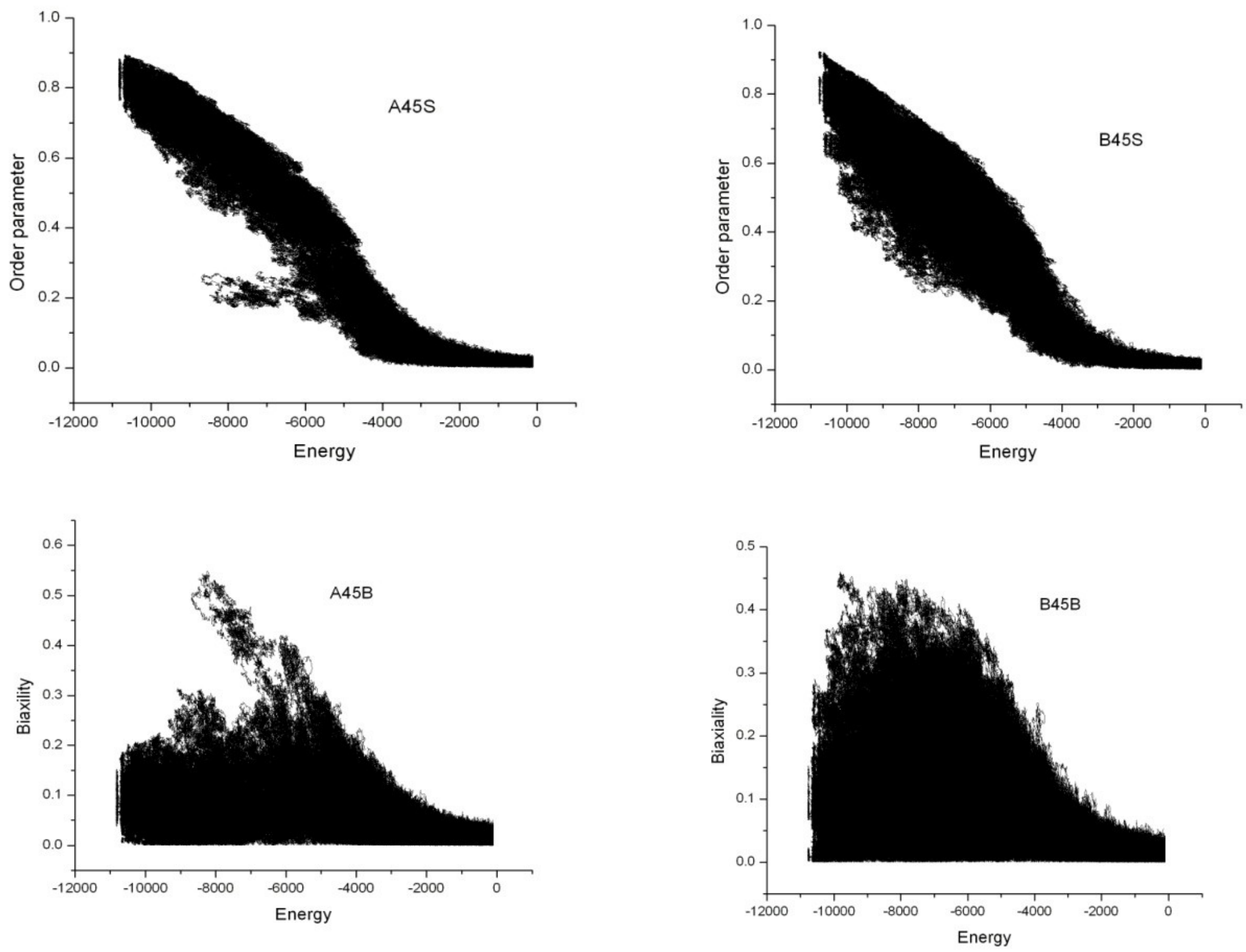

Figure 5: Uniaxial (top) and biaxial (bottom) order parameters for $\theta_D = 0^0$ (left) and $45^0$ (right) respectively for Model B.

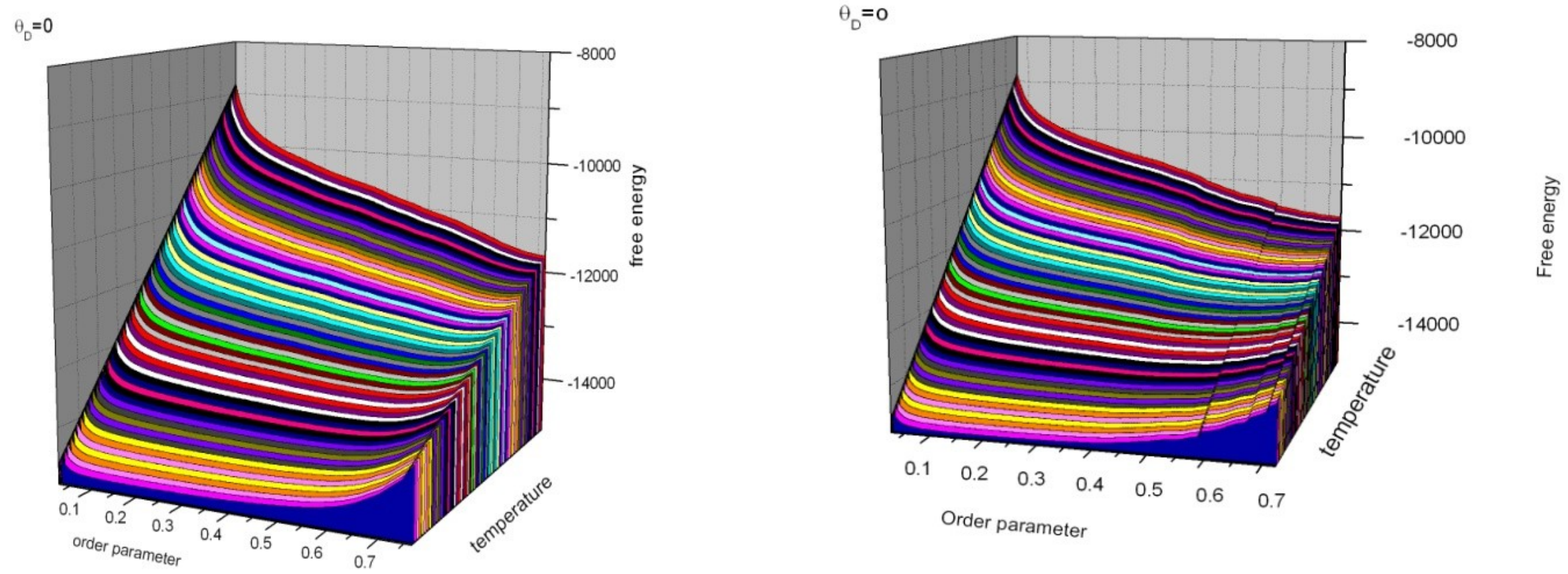

Figure 6: Free energy profiles at various temperatures for $\theta_D = 0^0$ for Models A (left) and B (right) respectively. Temperature increases as we go from left to right in the figures.

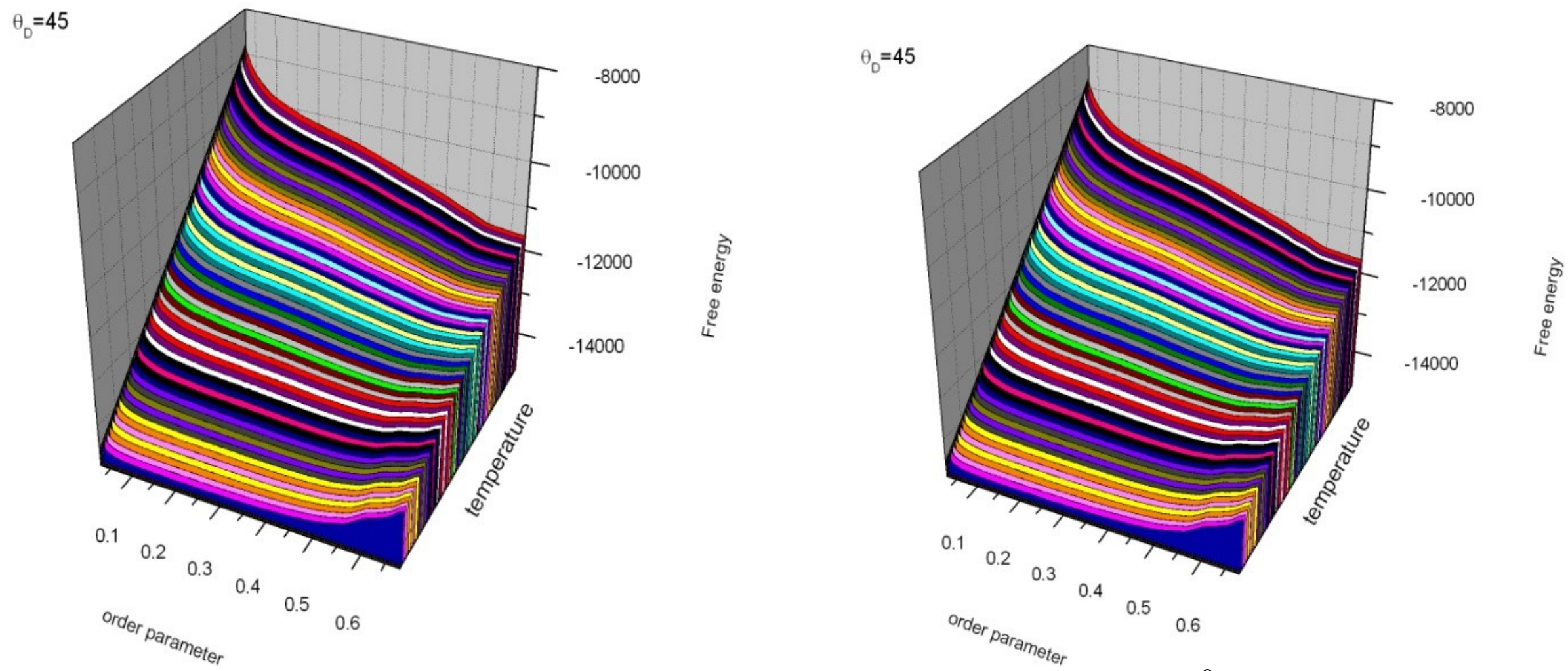

Figure 7: Free energy profiles at various temperatures for $\theta_D = 45^0$ for Models A (left) and B (right) respectively. Temperature increases as we go from left to right in the figures.

**III. Conclusions:**

In the present work we showed that non-Boltzmann Monte Carlo methods based on a simple lattice model, mimicking the boundary conditions of Model A, for example, show that the H phase expected of the model from considerations of a two dimensional system on grounds of continuum theory is an asymptotic limit of the chosen model at very low temperatures.

Secondly, the relative phase shift of the geometrical structure with respect to the chemical pattern has significant effect on the formation of director structures, as well as the apparent effective angle experienced at the lower substrate. Further, we observe an onset of biaxiality in the system as we increase the angle, $\theta_D$, which is an important consequence not realized in the earlier work on this system.

**Acknowledgements:**
The authors gratefully acknowledge the Centre for Modelling Simulation and design for providing computational resources and time for carrying out the present work. This work has been carried out under the DST project No. UH/CMSD/HPCF/2006-7.

**References:**

1. T. Rasing and I. Musevic, Surfaces *and Interfaces of Liquid Crystals,* 2004.
2. C. Uche, S. J. Elston and L. Parry-Jones, J. Phys. D: Appl. Phys. 38, 2283 (2005).
3. C. Tsakonas, et. al., Appl. Phys. Lett., 90, 111913 (2007).
4. S. Ladak, A. Davidson, et. al., J. Phys. D: Appl. Phys. 42, 085114 (2009).
5. R. Barberi, et. al., J. Appl. Phys. 84, 3, 1321 (1998).
6. J-H.Kim, et. al., Appl Phys. Lett. 78, 20, 3055 (2001).
7. G. Barbero, G. Skacej , A. L. Alexe-Ionescu and S. Zumer, Phys. Rev. E 60, 628 (1999).
8. C. V. Brown, M. J. Towler, V. C. Hui, and G. P. Bryan-Brown, Liq. Cryst. 27, 233 (2000).
9. C. V. Brown, L. A. Parry-jones, S. J. Elston, and S. J. Wilkins, Mol. Cryst. Liq. Cryst 410, 945 (2004).
10. L. A. Parry-Jones, E. G. Edwards, S. J. Elston, and C. V. Brown, Appl. Phys. Lett. 82, 1476 (2003); L. A. Parry-Jones, E. G. Edwards, and C. V. Brown, Mol.
11. Cryst. Liq. Cryst. 410, 955 (2004).
12. P. Patricio, M. M. Telo da Gamma, and S. Dietrich, Phys. Rev. Lett. 88, 245502 (2002).
13. X. Lu, Q. Lu, Z. Zhu, J. Yin, and Zongguang Wang, Chem. Phys. Lett. 377, 433 (2003).
14. D. H. Chung, T. Fukuda, et. al., J. Appl. Phys.92, 1841 (2002).
15. G. Barbero, T. Beic, A. L. Alex-Ionescu and R. Moldovan, J. Phys. II. 2, 2011 (1992).
16. T. Z. Qian and P. Sheng, Phys. Rev. Lett. 77, 4567 (1996); T. Z. Qian and P. Sheng, Phys. Rev. E. 55, 7111 (1997).
17. V. K. Gupta and N. L. Abbott, Science 276, 1533 (1997).
18. S. Kondrat and A. Poniewierski, Phys. Rev. E., 64, 031709 (2001).
19. A. Poniewierski and S. Kondrat, J. Mol. Liq. 112, 61 (2004).
20. S. Kondrat, A. Poniewierski and L. Harnau, Eur. Phys. J. E 10, 163 - 170 (2003).
21. L. Harnau, S. Kondrat and A. Poniewierski, Phys. Rev. E., 72, 011701 (2005).
22. T. J. Atherton and J. R. Sambles, Phys. Rev. E., 74, 022701 (2006).
23. L. Harnau, S. Kondrat and A. Poniewierski, Phys. Rev. E., 76, 051701 (2007).
24. F. C. Frank, Discuss. Faraday Soc. 25, 19 (1958).
25. P. G. de Gennes and J. Prost, The Physics of liquid crystals, 2nd ed., Clarendon, Oxford (1993).
26. A. Rapini and M. Papoular, J. Phys. (Paris), Colloq. 30, C4-54 (1959).
27. Jun-ichi Fukuda, Makoto Yoneya, and Hiroshi Yokoyama, Phys. Rev. E 79, 011705 (2009).

28. P. A. Lebwohl and G. Lasher, Phys. Rev. A 6, 426 (1972).
29. F. Barmes and D. J. Cleaver, Phys. Rev. E 69, 061705 (2004); F. Barmes and D. J. Cleaver, Chem. Phys. Lett. 425, 44 (2006).
30. F. Wang, and D. P. Landau, Phys. Rev. Lett. 86 2050 (2001); F. Wang and D. P. Landau, Phys. Rev. E 64 056101 (2001).
31. D. Jayasri, V. S. S. Sastry and K. P. N. Murthy, Phys. Rev. E 72, 036702 (2005); P. Poulain, F. Calvo, et. al., Phys. Rev. E. 73 056704 (2006).
32. C. Zhou, T. C. Schulthess, S. Torbrugge, and D. P. Landau, Phys. Rev. Lett. 96, 120201 (2006).